\documentclass[twocolumn,twoside,slac_two]{revtex4}
\usepackage{graphicx}
\usepackage{fancyhdr}
\def\la{\mathrel{\mathchoice 
{\vcenter{\offinterlineskip\halign{\hfil$\displaystyle##$\hfil\cr<\cr\sim\cr}}}
{\vcenter{\offinterlineskip\halign{\hfil$\textstyle##$\hfil\cr<\cr\sim\cr}}}
{\vcenter{\offinterlineskip\halign{\hfil$\scriptstyle##$\hfil\cr<\cr\sim\cr}}}
{\vcenter{\offinterlineskip\halign{\hfil$\scriptscriptstyle##$\hfil\cr<\cr\sim\cr}}}}}
\def\ga{\mathrel{\mathchoice 
{\vcenter{\offinterlineskip\halign{\hfil$\displaystyle##$\hfil\cr>\cr\sim\cr}}}
{\vcenter{\offinterlineskip\halign{\hfil$\textstyle##$\hfil\cr>\cr\sim\cr}}}
{\vcenter{\offinterlineskip\halign{\hfil$\scriptstyle##$\hfil\cr>\cr\sim\cr}}}
{\vcenter{\offinterlineskip\halign{\hfil$\scriptscriptstyle##$\hfil\cr>\cr\sim\cr}}}}}
\begin{document}

\title{Extensive near infrared monitoring of millimeter-wave / gamma-ray
bright blazars}

%

\author{Alberto Carrami\~nana, Luis Carrasco, Alicia Porras, Elsa Recillas} 
\affiliation{INAOE, Luis Enrique Erro 1, Tonantzintla, Puebla 72840, M\'exico}
\begin{abstract}
We established a sample of millimeter-wave and $\gamma$-ray bright active galactic
nuclei matching the {\em WMAP} catalog with the EGRET catalog, highest energy
photons and the {\em Fermi} bright source list.  We have monitored over 80 of these 
objects in the near infrared, obtaining over 2000 JHKs data points directly comparable 
with {\em Fermi} data.  We present examples of correlated near infrared and $\gamma$-ray 
activity of known blazars and recently identified sources.
\end{abstract}

\maketitle

\thispagestyle{fancy}

\section{mm-wave $\gamma$-ray bright blazars}
{\em CGRO}-EGRET established the predominance of flat spectrum radio quasars 
(FSRQ) as extragalactic sources of high-energy $\gamma$-rays~\cite{3egcat}. 
These are believed to be powered by supermassive black holes ejecting large
amounts of material along relativistic jets perpendicular to accretion disks and
producing high energy particles in these jets. Observations at the highest photon 
energies shown a predominance of Bl Lac objects over FSRQ at $E\sim 1~\rm TeV$. 
Although these observations may be biased by the horizon limitation due to pair absorption 
with extragalactic background light (EBL), the {\em Fermi} $\gamma$-ray Space Telescope, 
more sensitive and more responsive to GeV photons than EGRET, reported a larger 
fraction of Bl Lac objects among blazars in its bright source list release 
(0FGL)~\cite{fermi-0fgl,fermi-0fgl-agn}. The wider spectral coverage of {\em Fermi} is 
allowing detailed studies of the intrinsic properties of various types of active galactic 
nuclei (AGN), cosmic evolution and their use as EBL tracers~\cite{dermer2007}.

The extrapolation of FSRQ to high radio frequencies makes the coincidence of 
foreground {\em WMAP} sources with $\gamma$-ray blazars expectable. Furthermore, 
it is clear than most of the 390 {\em WMAP} sources are AGNs~\cite{wmap-cat}. These 
emitters are dominated in the mm-wave band by the same synchrotron component 
observed at lower frequencies and can be expected to emit photons of very high energy, 
related to the hardest part of the synchrotron emission. {\em WMAP} sources have fluxes 
above 0.1~Jy at frequencies $\ga 40~\rm GHz$, making them suitable positional references 
for large millimeter telescopes. Known common {\em WMAP}/EGRET sources, found up to 
$z\ga 2.3$, are relatively nearby analogs of more distant blazars detectable in principle by
{\em Fermi} up to $z\ga 7$ and by the Large Millimeter Telescope (LMT~\cite{gtm}) 
at $z\ga 30$, if such an object could have existed at such an early phase of the Universe.  
The combination of {\em Fermi} and LMT data promises the availability of a sample of 
thousands of objects up to the highest redshifts ideally suited to study radio loud AGN 
evolution and the connection of the EBL with the star formation history of the Universe.

\begin{table}[b]
\begin{center}
\caption{{\em WMAP} and EGRET source matches by type.\\ 
($\dagger$ = excludes P, G and S types~\cite{3egcat}).\label{g2wmask}}
\begin{tabular}{|c|c|c|c|c|}
 \hline\hline \textbf{Source}  & \textbf{3EG} &  \textbf{{\em WMAP}} &
\multicolumn{2}{c|}{\textbf{Matches}}  \\
 \textbf{Type}  & \textbf{catalog} & \textbf{mask} & \textbf{Observed} & 
 \textbf{Expected} \\
 \hline  A & 67 & 54 & 40 & 6.6 \\
 \hline a & 28 & 23 & 17 & 3.3 \\
 \hline U & 170 & 61 & 11 & 10.0 \\
 \hline Total & 265$^{\dagger}$ & 138 & 68 & 20 \\
 \hline\hline
 \end{tabular} \end{center} \end{table}

\begin{table*}[t]
\begin{center}
\caption{{em WMAP} sources in the 3EG catalog and 0FGL bright source list -- Part I.\\ 
$D$ denotes the integer part of the distance in standard deviations.\label{muestra1}}
\begin{tabular}{|l|c|c|c|c|l|c|c|c|c|}
\hline \textbf{WMAP} & \textbf{3EG / 0FGL} & \textbf{$D$} & \textbf{Counterpart} &&
           \textbf{WMAP} & \textbf{3EG / 0FGL} & \textbf{$D$} & \textbf{Counterpart}  \\
\hline 
\hline J0050--0649 & 0FGL J0051.1--0647 & [0] & PKS 0048--071 & ~ & 
            J0334--4007  & 0FGL  J0334.1--4006 & [0] & PKS 0332--403  \\  
\hline J0051--0927 & 0FGL J0050.5--0928 & [0] & PKS 0048--097  && 
            J0339--0143  &3EG J0340--0201& [0]   & CTA 026  \\  
\hline J0108+0135  & 3EG J0118+0248   & [2]   &   4C 01.02   && 
           J0407--3825  &0FGL  J0407.6--3829& [0]  & PKS 0405--385 \\
\hline J0132--1653  &  3EG J0130--1758   &[1]  & QSO B0130--171   &&  
            J0416--2051 &3EG  J0412--1853 & [1]   & (QSO B0413--21)?? \\   
\hline J0137+4753  & 0FGL J0137.1+4751 &  [0]   & QSO B0133+47    && 
           J0423--0120   & 0FGL J0423.1--0112 & [0]  & QSO~B0420--015 \\  
          &&&&&                & 3EG J0422--0102 & [0]  & \\  
\hline J0204+1513  & 3EG J0204+1458 & [0] & 4C +15.05   &&  
          J0428--3757 & 0FGL J0428.7--3755 & [0]  & PKS 0426--380 \\ 
\hline J0205--1704 & 0FGL J0204.8--1704 & [0]  & PKS 0202--17   &&
           J0442--0017 & 3EG  J0442--0033& [0]   & QSO B0440--004 \\ 
\hline J0210--5100  & 0FGL J0210.8--5100 & [0]  & QSO B0208--5115   && 
           J0455--4617  & 3EG  J0458--4635& [0] & 0454--463    \\
           & 3EG J0210--5055  & [0]  &   && &&&   \\
\hline J0218+0138  & 0FGL J0217.8+0146 & [1]   & PKS 0215+015   &&
           J0456--2322  & 0FGL J0457.1--2325 & [0]  & QSO B0454--234  \\ 
                          &&&&&& 3EG J0456--2338 & [0]  &  \\
\hline J0220+3558 & 0FGL J0220.9+3607 & [0] & B2 0218+35   &&  
         J0501--0159 & 3EG  J0500--0159 & [0]  & 4C --02.19  \\
\hline J0223+4303  &  0FGL J0222.6+4302 & [1]   &  3C 66A  && 
           J0506--6108  & 3EG J0512--6150 & [1]  & 0506--612?   \\ 
                        &  3EG J0222+4253  & [0]   &    && &&&       \\
\hline J0237+2848  & 0FGL  J0238.4+2855 & [0] & 4C +28.07    && 
           J0523--3627  & 3EG J0530--3626 &  [2]  & QSO J0522--3627   \\
                         &   3EG J0239+2815 & [1] &   && &&&   \\
\hline J0238+1637 & 0FGL J0238.6+1636 &  [0]  & AO 0235+164   &&
          J0538--4405  &  0FGL J0538.8--4403 &    [0]   & PKS 0537--441      \\
                        & 3EG J0237+1635 &    [1]     & QSO B0235+16   &&   
                        &  3EG J0540--4402 &  [0]   &   \\
\hline J0319+4131  & 0FGL J0320.0+4131 & [0]  & NGC 1275    &&  
           J0539--2844 & 3EG  J0531--2940 &  [2]  & (QSO~J0539--2839)??  \\ 
\hline\hline
\end{tabular}\end{center}\end{table*}

\section{Matching {\em WMAP} with $\gamma$-ray sources}
As a by product of the study of the cosmic microwave background, the {\em Wilkinson 
Microwave Anisotropy Probe (WMAP)} produced a catalog of 390 bright sources detected 
at frequencies between 23 and 94 GHz, outside a mask defined by the mm-wave emission 
from the Galactic plane~\cite{wmap-cat}. These objects constitute a fairly homogeneous 
sample suitable for comparison with all-sky $\gamma$-ray catalogs. 

We compared the {\em WMAP} foreground source catalog~\cite{wmap-cat} with:
(1) the Third EGRET (3EG) catalog of high-energy $\gamma$-ray sources~\cite{3egcat};
(2) the $E>10~\rm GeV$ EGRET photons compiled by~\cite{egret-vhe}; (3) the list of bright 
sources detected with signal-to-noise ($\sqrt{TS}$) larger than 10 by the {\em Fermi} 
$\gamma$-ray Space Telescope in its first three months of operations~\cite{fermi-0fgl}.
The positional uncertainty of {\em WMAP} equals $4^\prime$, similar to that of {\em Fermi},
but much better than that of EGRET. Preferring to accept false positives than to reject 
real coincidences, we used a relaxed matching criterion of 2.5 times the combined positional 
accuracy, $(\sigma_{wmap}^{2}+\sigma_{\gamma}^{2})^{1/2}$, where $\sigma_{\gamma}$ 
refers to 3EG ($\sim 1^{\circ}$), EGRET-VHE ($0.5^\circ$) or 0FGL ($2-10^{\prime}$). 
For each comparison we estimated the expected number of random coincidences quantifying 
the solid angle covered by the $\gamma$-ray sources (or events) outside the {\em WMAP} 
Galactic mask. For each {\em WMAP}/$\gamma$ pair we blindly listed potential radio, optical 
and X-ray counterparts from SIMBAD / Vizier. 
The results are listed in tables~\ref{muestra1} and~\ref{muestra4}.

\subsection{Matching {\em WMAP} with EGRET}
The comparison between the {\em WMAP} and 3EG catalogs produced 69 matches out of 
the 390 {\em WMAP} and 138 EGRET sources outside the {\em WMAP} Galactic mask. 
The number expected randomly is 20; the probability of having 69 matches among the
390 {\em WMAP} sources is $P\la 10^{-17}$, indicating that most -but not all- the matches 
are real. 

When accounting for the source type we count 40 matches out of the 54 high confidence 
blazar association, labelled A in the 3EG catalog, compared to 6.6 expected randomly. 
The low confidence "a" associations have 17 matches out of 23 tries and 3.3 expected 
chance coincidences. On the other hand we have 11 matches among the 61 unidentified
sources out of the {\em WMAP} mask, expecting 10.0 by chance. On statistical grounds, 
we can confirm  the physical association of foreground {\em WMAP} sources with 3EG 
blazars, accounted by both "A" and "a" classes, but not with unidentified EGRET sources 
(table~\ref{g2wmask}).

We also compared the {\em WMAP} positions with the list of very high energy (VHE) 
photons, $E>10\,\rm GeV$~\cite{egret-vhe}: 510 out of the 1506 VHE photons
are outside the {\em WMAP} Galactic mask. The combined positional accuracy of 
VHE photons and {\em WMAP} sources is $30.2^\prime$. We obtained coincidences
between 33 VHE photons and 29 {\em WMAP} sources as follows:\\
-- 20 {\em WMAP} sources coincide with a single isolated VHE photon;\\
--   4 {\em WMAP} sources coincide with a single VHE photon and a 3EG,
with no 0FGL counterpart;\\
--   1 {\em WMAP} source (WMAP J1408--0749) coincides with a 3EG source
(3EG J1409--0745) and 4 VHE photons (VHE 494, 498, 1058 and 1061) -- but
with no 0FGL counterpart:\\
-- 3 {\em WMAP} sources coincide with VHE photons, 3EG and  
0FGL sources;  of these WMAP~J0210-5100 has two VHE photons;\\
-- 1 {\em WMAP} source (WMAP~J0137+4753) coincides with a VHE photon
and a 0FGLsource (0FGL~0137.1+4751), with no 3EG counterpart.\\
We note that given the number and error box sizes of the VHE photons, we expect
30 random matches under our 2.5$\sigma$ criterion.  Most or all of the single
{\em WMAP}-VHE coincidences are likely to be spurious. We note the case of
WMAP~J0137+4753, which belonged to the {\em WMAP}\& VHE only category prior 
to the publication of the 0FGL and turned out to be a real $\gamma$-ray source.

\subsection{Matching the {\em WMAP} catalog with the {\em Fermi} bright source list}
The {\em Fermi} bright source list (0FGL), made public in February 2009, consists 
of 205 bright $\gamma$-ray sources detected with significance $\sqrt{TS}> 10$ 
in the first three months of observations~\cite{fermi-0fgl}. Of these, 121 are of the
AGN class, mostly blazars~\cite{fermi-0fgl-agn}. 
The 0FGL list has 122 objects outside the {\em WMAP} Galactic mask, which we
compared with the respective {\em WMAP} and {\em Fermi} positions. The improved 
positional accuracy of {\em Fermi}, in the $4^{\prime}-10^{\prime}$ range,  results in 
only 0.82 spurious coincidences expected. We 
found 54 matches between  {\em WMAP} and the 0FGL, 25 of which  are common with 
EGRET sources and 29 are independent.

\subsection{Sample overview}
The sample is presented in tables~\ref{muestra1} and~\ref{muestra4}. The $D$ column 
expresses the distance between the {\em WMAP} and the $\gamma$-ray event in terms 
of the integer part of the combined positional accuracy; associations with [0] have 
intersecting error countours and are more likely to be real than those with [2], separated 
by $\geq 2\sigma$.  Most of the sources have a suitable radio, optical and/or X-ray counterpart, 
often in the intersection of the {\em WMAP}/$\gamma$-ray error boxes. In a few cases the 
candidate counterpart is not unique and the one displayed is a subjective election. 
Counterparts in parenthesis are tentative. 
We note the following:
\begin{itemize}
\item WMAP J0051--0927 :: 0FGL J0050.5--0928 has a preferred association with 
the Bl Lac object PKS 0048--071 
= PHL 856; the radiogalaxy FIRST J005051.9--092529 is an alternative.
\item WMAP J0237+2848 :: 0FGL J0238.4+2855 :: 3EG J0239+2815.
This is listed as a low confidence {\em Fermi} association with 4C +28.07
in~\cite{fermi-0fgl}. The {\em WMAP} and {\em Fermi} boxes match, both containing  
4C +28.07. All are somewhat outside the EGRET box. 
\item WMAP~J0319+4131 :: 0FGL J0320.0+4131 is identified with the radiogalaxy 
NGC 1275. We note that the radiogalaxy Cen~A is excluded of this study by 
the {\em WMAP} Galactic mask.
\item WMAP J0423--0120 :: 0FGL J0423.1--0112 :: 3EG J0422--0102. 
This is a low confidence {\em Fermi} association with PKS~0420--014, which is 
at the center of the {\em WMAP} circle.
\item WMAP J0909+0119 :: 0FGL J0909.7+0145. The 0FGL error radius is 
$17^{\prime}$, the {\em WMAP} and 0FGL positions differing by $25^\prime$.
PKS 0907+022 is a low confidence {\em Fermi} association not compatible with 
{\em WMAP}. 4C+01.24 (PKS = QSO B0906+015) is inside the {\em WMAP} box and 
marginally compatible with the {\em Fermi} source. 
\item WMAP~J1517--2421 has two possible EGRET counterparts but only one 
(3EG J1517--2538) is compatible with the 0FGL source.
\item WMAP~J1642+3948 is $19^\prime$ away from 0FGL~J1641.4+3939, which has
an positional error of $9.5^\prime$, making the association tentative. There are suitable
candidates for both sources, with 3C345 being positionally the best {\em common} 
counterpart, as discussed in \S\ref{3c345-id}.
While there is no 3EG source association, we  note the revised EGRET counterpart 
EGR~J1642+3940~\cite{egrcat}. 
\end{itemize}

\section{Near infrared monitoring}
On August 2007, we started a dedicated monitoring program
with up to 60 nights per semester awarded on the 2.1m telescope of the Observatorio 
Astrof\'{\i}sico Guillermo Haro (OAGH), in Cananea, Sonora, Mexico 
(lat=$+31.05$, long=$-110.38$). The program consists of optical photometry 
(BRVI), low resolution spectroscopy (4000 - 7500 \AA) and near infrared JHKs imaging 
of the sample above, with the addition of high priority {\em GLAST/Fermi} sources and 
targets notified via the multi-wavelength {\em Fermi} group. The current study is to lead 
to programs of follow-up and identification of $\gamma$-ray sources using near infrared, 
optical and mm-wave facilities. 

\begin{figure}[t]
\centering\includegraphics[width=\hsize]{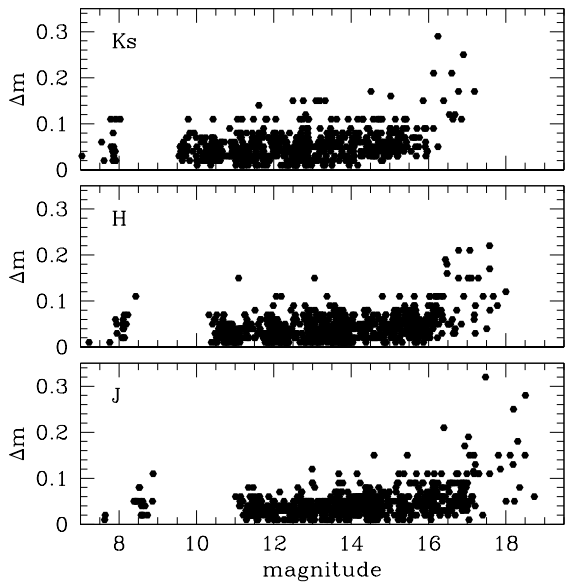}
\caption{CANICA photometric measurements of mm-wave/$\gamma$-ray sources. 
We show only detections.\label{fotometria}}
\end{figure}

Particularly  successful has been the JHKs photometric survey using the CAnanea 
Near Infrared CAmera (CANICA).  CANICA is equiped with a Rockwell $1024\times
1024$ pixel Hawaii infrared detector working at 75.4~K with standard near infrared filters.
The scale plate is 0.32$^{\prime\prime}$/pixel. Observations are usually carried out in 
series of dithered frames in each filter. Datasets are coadded after correcting for bias and
flat-fielding using IRAF based macros. Figure~\ref{fotometria} shows the photometric 
magnitudes measured with CANICA, with detection thresholds around magnitudes 19, 18 
and 17 for J, H and Ks respectively, i.e. about two magnitudes fainter than 2MASS.

\section{Correlated infrared/$\gamma$-ray activity}
We have found correlated infrared and $\gamma$-ray correlations with little
or no evidence for time delays in our sample.  We show here joint CANICA 
H band (1.6$\mu\rm m \Rightarrow 0.76 eV$) and {\em Fermi} ($1-300 \,\rm GeV$) light 
curves, normalizing the maximum $\gamma$-ray flux to the maximum H-band flux and 
setting both zeros at the same level. We are currently studying 
the infrared color behavior of our sample during flares and constraining possible
delays in the timing of near infrared flares relative to the $\gamma$-ray ones.
Our current results on  3C 454.3 are presented in~\cite{g2mm-3c454}.

\subsection{PKS 0235+164}
\begin{figure}[t]
\centering\includegraphics[width=\hsize]{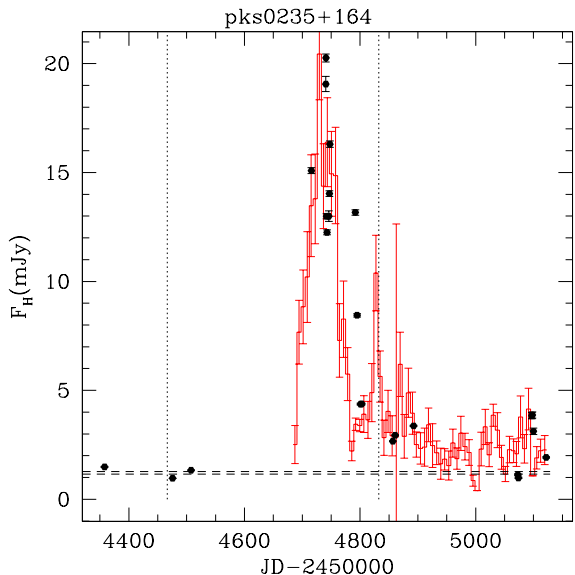}
\caption{Joint CANICA - {\em Fermi} light curve of PKS 0235+164. CANICA is given
by the (black) points while {\em Fermi} fluxes form the (red) histogram with error bars.
The discontinuous horizontal line marks the 2MASS flux. The dotted vertical lines 
indicate the change of year. \label{h2f-pks0235+164}}
\end{figure}
PKS 0235+164 is positionally coincident with WMAP~J0238+1637, mm-wave
source coincident with 3EG~J0237+1635 = 0FGL~J0238.6+1636. The {\em WMAP}
source is labelled as probable variable. Figure~\ref{h2f-pks0235+164} shows the
joint H band and 1-300~GeV fluxes scaled.
The near infrared flux at the end of 2008 and beginning of 2009 matched the 
2MASS values. On JD2454715 we found PKS~0235+164 almost three magnitudes
brighter, prior to peaking 25 days later - in coincidence with the {\em Fermi} flare.

\subsection{PKS 0454--234}
\begin{figure}[t]
\centering\includegraphics[width=\hsize]{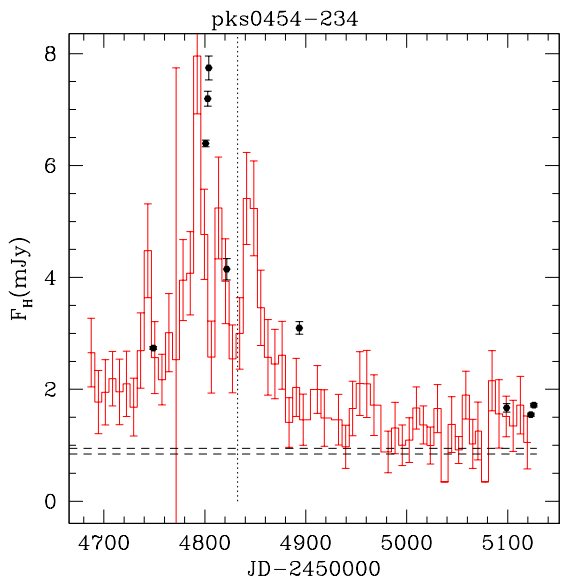}
\caption{Joint CANICA - {\em Fermi} light curve of PKS 0454--234. The discontinuous
horizontal line marks the 2MASS flux. The dotted vertical lines indicate the change
of year.\label{h2f-pks0454-234}}  
\end{figure}

The {\em WMAP} and {\em Fermi} error circles in this region of the sky intersect close to 
the center of the 3EG~J0456--2338 error box. PKS 0454--234 is within the intersection,
in a neat positional match. We started monitoring on JD 2454748, when the near
infrared flux was  0.85 magnitudes above the 2MASS reference value. The source
flared by a factor of nearly 3 in the next 50 days to drop to a half 18 days later.
The near infrared peak seems delayed by about a week relative to the 1-300 GeV 
maximum. The AGN has been in relative quiescence since (fig.~\ref{h2f-pks0454-234}).

\subsection{PKS 1510--089}
PKS 1510--89 is the high confidence counterpart of 3EG J1512--0849, WMAP 
J1512--0904 and 0FGL~J1512.7--0905, with an excellent positional match
between all data. We started monitoring PKS 1510--89 in early 2008, a few months
prior to the {\em Fermi} launch, when the flux fluctuated around the 2MASS 
reference value. We caught a simultaneous infrared - $\gamma$-ray flare 
on JD 2454924 and the subsequent decay a month later.
\begin{figure}[t]
\centering\includegraphics[width=\hsize]{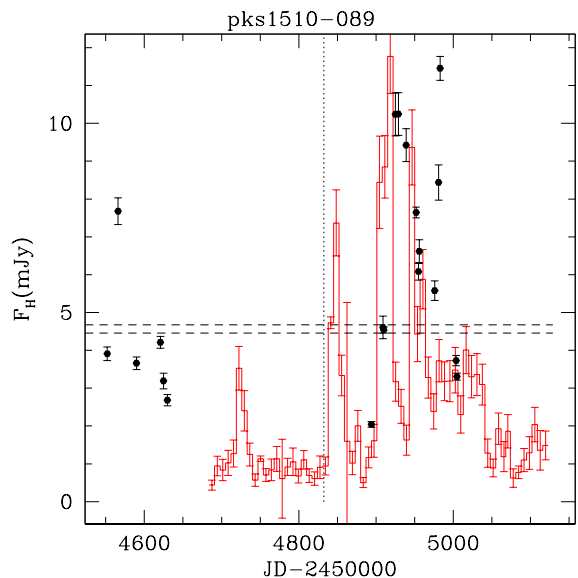}
\caption{CANICA - {\em Fermi} light curve of PKS 1510--089. The discontinuous
horizontal line marks the 2MASS flux. The dotted vertical lines indicate the change
of year.\label{h2f-pks1510-089}}
\end{figure}

\section{Source identifications}
\subsection{The identification of QSO B0133+476}
QSO B0133+476 is a rather active quasar first catalogued originally as DA55
in the Dominion DA 1420~MHz survey~\cite{da1968}.  Also known as Mis V1436, 
archival optical data exists since at least 1953, when it was around magnitude R=18.7. 
This object has been monitored by other groups since 2007, when it was found to be 
4.5 magnitudes brighter. Its optical variability is well documented in~\cite{misao}.

\begin{figure}[t]
\centering\includegraphics[width=\hsize]{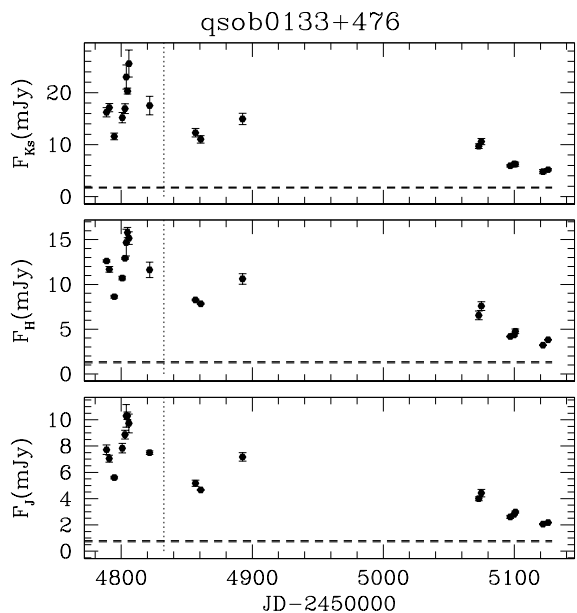}
\caption{CANICA light curve of QSO B0133+47. The flaring at the end of 2008 is
clearly visible, at a factor of 10 above the 2MASS flux levels indicated by the dashed
horizontal line.\label{go-qso0133+476}}
\end{figure}

This object has no EGRET counterpart, the only evidence for $\gamma$-ray emission
prior to the {\em Fermi} launch being its marginal closeness to a $85(\pm 38)\rm GeV$ photon. 
The unprovable association of this single photon with QSO B0133+476 is of interest 
as its energy is close to the pair absorption EBL limit for the redshift of the object, 
$z=0.859$. If true, this association would indicate the potential of this source for testing 
the EBL horizon using {\em Fermi}, specially under phases of high emission activity. 
Our report of near infrared flaring from data taken between JD 2454788  and 
2454795~\cite{qso0133-nir} was followed by the {\em Fermi} detection report, at a flux 
level above the EGRET limiting sensitivity~\cite{qso0133-gam}. The infrared light curve 
is shown in figure~\ref{go-qso0133+476}. The object was caught undergoing a rapid flare 
which has declined relatively slowly during 2009, but always at levels above 2MASS.
\begin{figure}[t]
\centering\includegraphics[width=0.49\hsize]{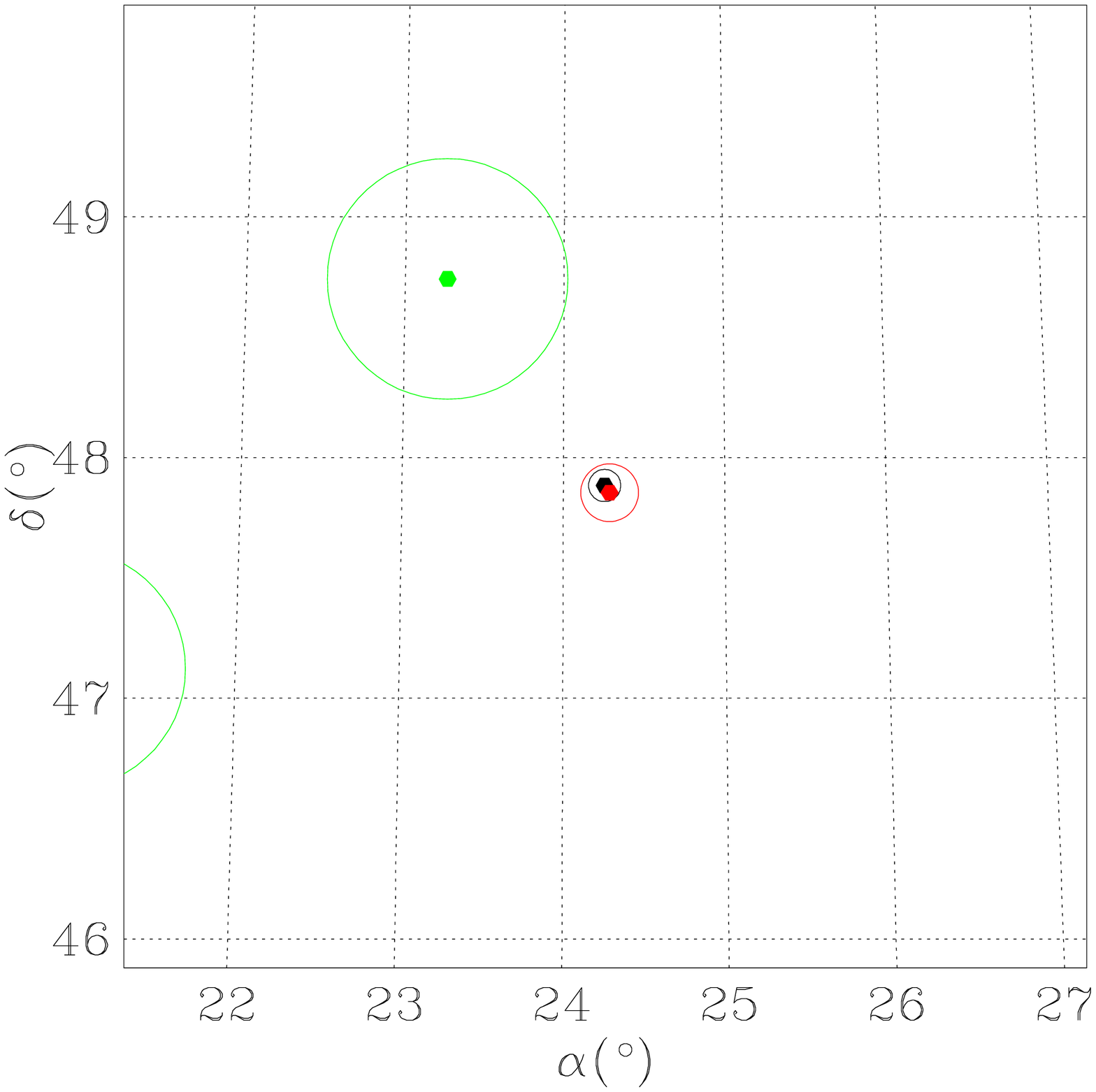}
                  \includegraphics[width=0.49\hsize]{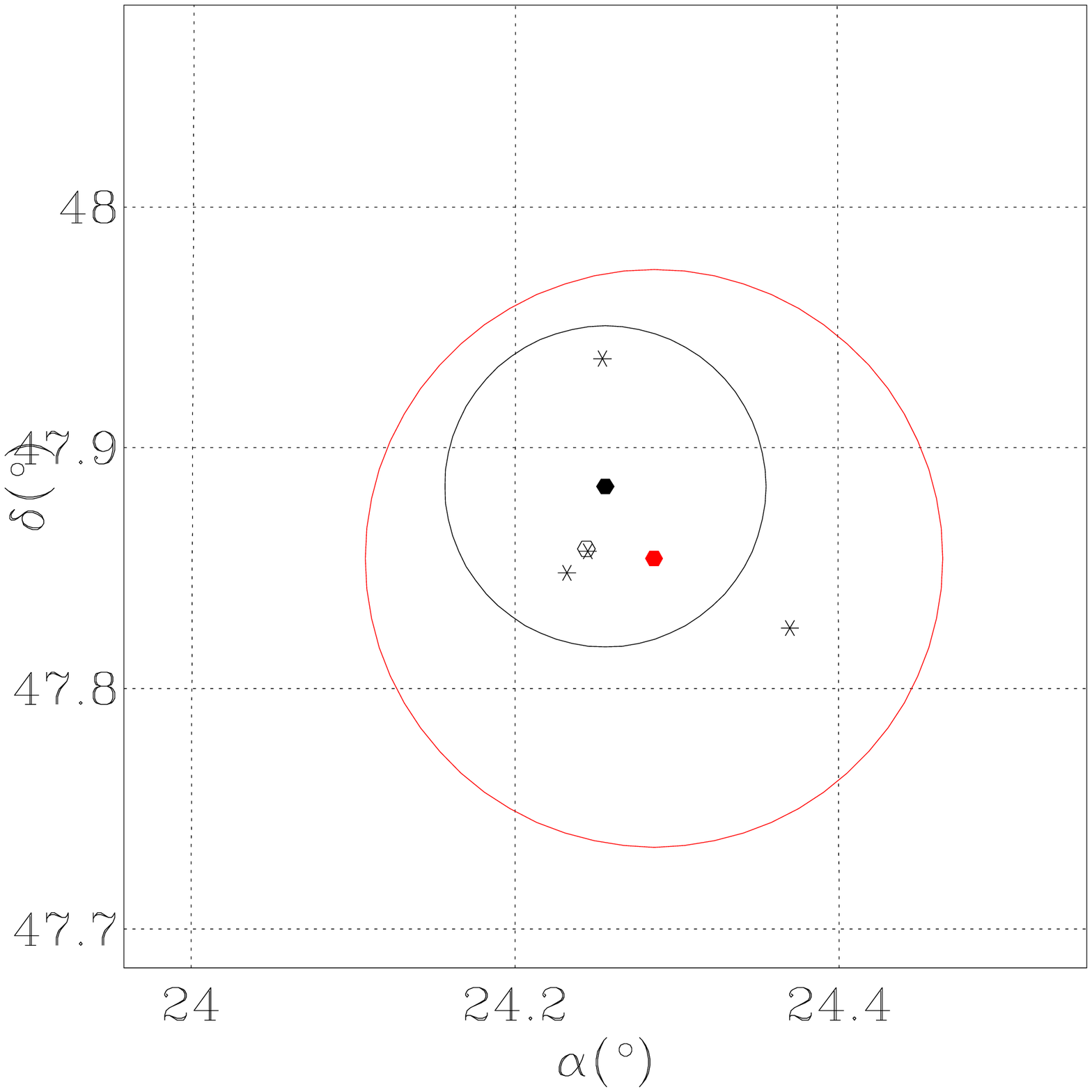}
\caption{Positional circumstance of QSO B0133+476. {\em Left:} the VHE photon error 
circle is shown at $\alpha\sim 23^{\circ}$, $\delta\sim 48.6^{\circ}$, with the almost concentric 
{\em WMAP} and {\em Fermi} circles in the middle. {\em Right:} a zoom panel showing 
the  $4^\prime$ {\em WMAP} circle contained in the  $7^\prime$  {\em Fermi} position. 
QSOB0133+476 is indicated by the open hexagon and an asterix.\label{j0137}}
\end{figure}

\subsection{QSO J0808--0751}
The EGRET source  3EG J0812--0646 was not in the high priority {\em GLAST/Fermi}
list, with the association with QSO~J0808--0751 been only tentative ($D=2$). 
WMAP~J0808--0750 is somewhat displaced from the 3EG location, but contains
this $z=1.84$ QSO.  The rapid flare observed simultaneously in the near infrared and 
by {\em Fermi} of this source, not in the 0FGL, confirms the identity of the QSO as the 
$\gamma$-ray source. We missed a second flare occurring while the source was on the 
daylight. The flux increase in the near infrared reached a factor of ten within 50 days. 
\begin{figure}[t]
\centering\includegraphics[width=0.98\hsize]{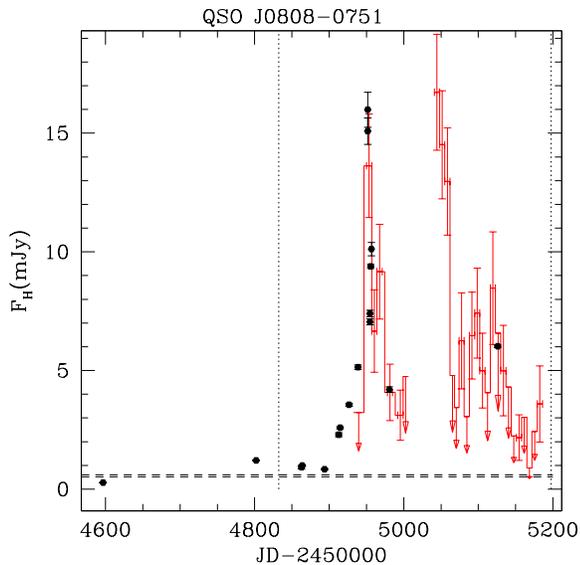}
\caption{CANICA - {\em Fermi} QSO J0808--0751 light curve. The dots indicate the H 
band fluxes in mJy, while the histogram and upper limit arrows indicate the
1-300~GeV fluxes and limits in scaled units. The discontinuous horizontal line marks 
the 2MASS flux. The dotted vertical lines indicate the change of  year.
\label{h2f-qsoj0808-0751}}
\end{figure}

\subsection{The identification of 0FGL~J1641.4+3939 with 3C 345 \label{3c345-id}}
\begin{figure}[t]
\vspace*{1mm}
\centering\includegraphics[width=0.49\hsize]{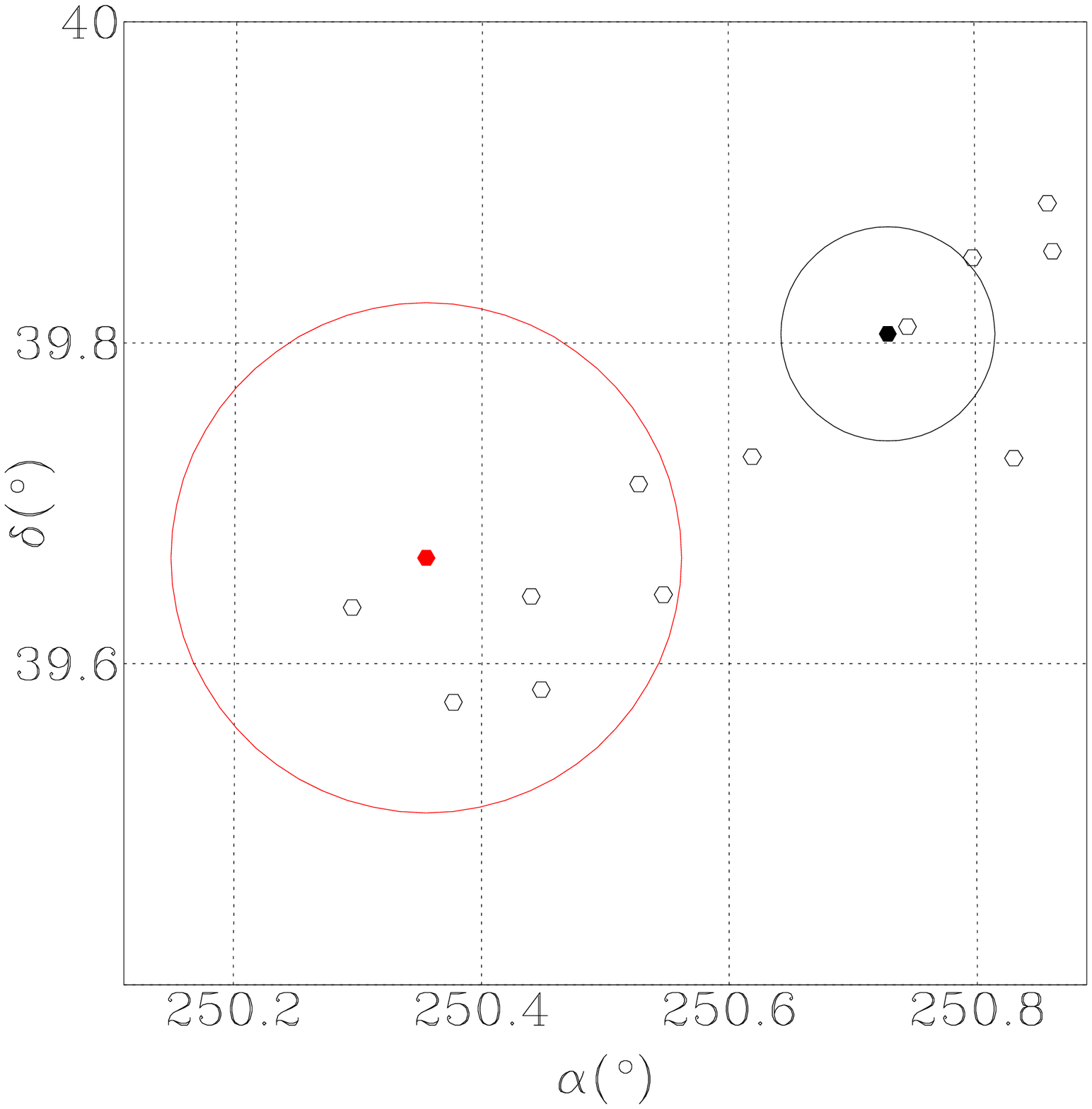}
                  \includegraphics[width=0.49\hsize]{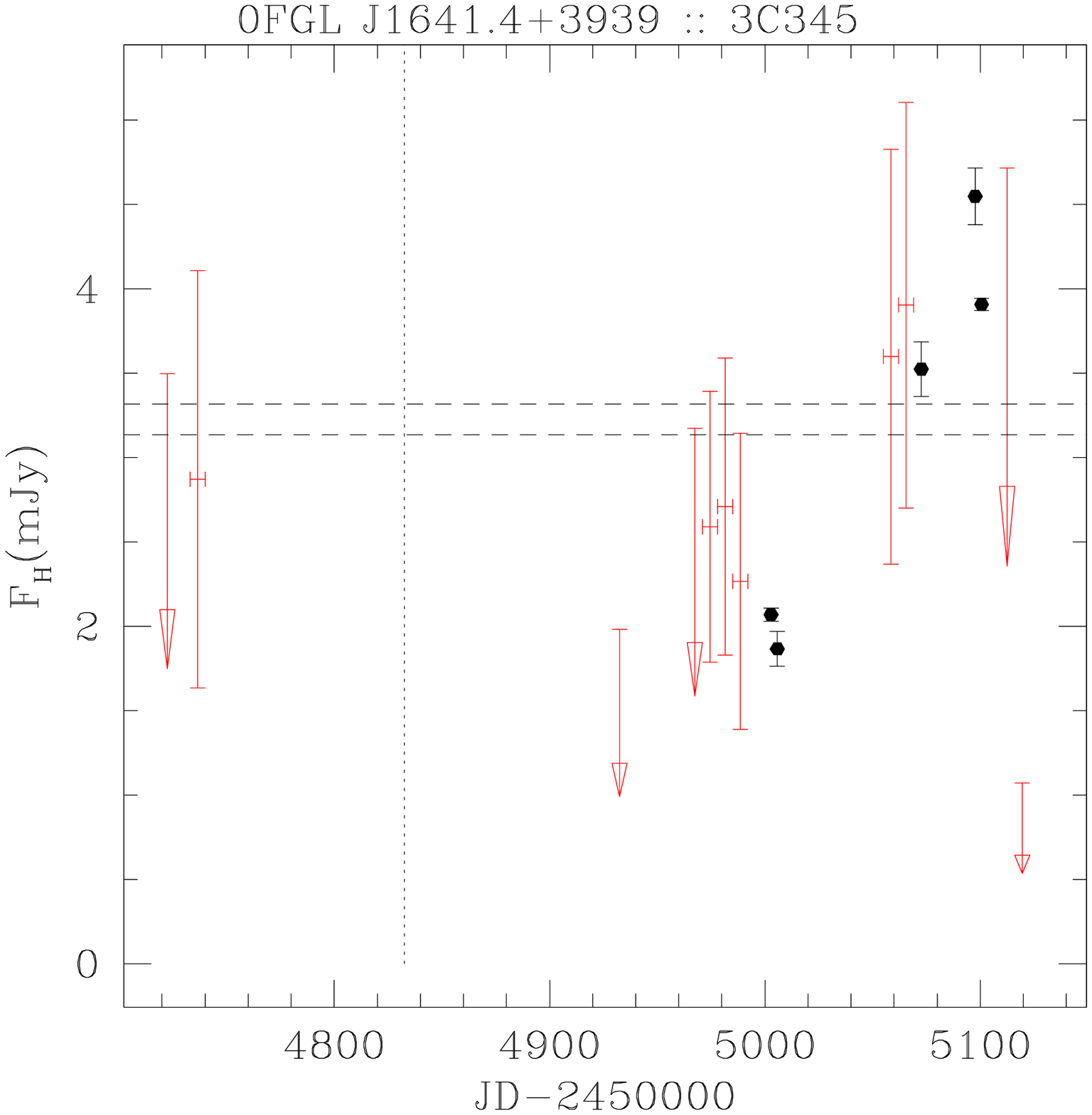}
\caption{{\em Left:} position of 0FGL~J1641.4+3939 (larger circle) and WMAP~J1642+3948 
(smaller circle) with a few potential counterparts (AGN class - open hexagons). 
{\em Right:} CANICA H band light curve of 3C345 compared with the (1-300 GeV) fluxes 
from 0FGL~J1641.4+3939 by {\em Fermi}.  The discontinuous horizontal lines mark the 2MASS 
flux and error. The dotted vertical line indicates the change of  year.\label{h2f-3c345}}
\end{figure}

The angular distance between WMAP~J1642+3948 and 0FGL~J1641.4+3939 is 
$19.2^\prime = 1.9\sigma$ times the combined positional uncertainty, the 
association between both objects being tentative only. This is a rather populated
region of the sky, where several potential counterparts can be found for each object:\\
- candidate counertparts of 0FGL~J1641.4+3939 include 3C 345, QSO B1641.5+3956, 
QSO B1641.6+3949, QSO B1640+398, QSO B1641.8+3956, QSO B1641+3958, 
QSO B1640.5+3944, QSO B1640+396, three more QSOs, a Seyfert 1 and a radiogalaxy 
from SDSS. \\
- potential counterparts for WMAP~J1642+3948 are GB6 J1642+3948 = 3C 345, 
FIRST J164304.3+394836, QSO B1641+3949, ...\\
3C~345 has a lower $\chi^{2}$ relative to the combined {\em WMAP} and {\em Fermi} 
positions (fig.~\ref{h2f-3c345}), being the best candidate under the assumption of a common 
association. Even though more data are desirable, the simultaneous flux changes measured 
provide evidence for the physical association between 0FGL~1641.4+3939 and 3C345.

\section{Summary}
We have selected  a sample of mm-wave/$\gamma$-ray bright blazars from the
{\em WMAP}, 3EG catalogs and 0FGL list. We have monitored these in the near
infrared finding correlated variability, which has also allowed the identification or 
confirmation of some of these objects.

%
%
%
%
%

%


\bigskip 
\begin{acknowledgments}
We acknowledge the use of 2MASS, WMAP, Fermi, SIMBAD, Vizier and POSS
databases. We appreciate the support of the technical staff at the Observatorio
Astrof\'{\i}sico Guillermo Haro.

\end{acknowledgments}
\bigskip 

\begin{thebibliography}{9}   
\bibitem{3egcat} Hartman, R.C., et al. 1999, ApJS 123, 79.
\bibitem{fermi-0fgl} Abdo, A.A., et al. 2009a, ApJS 183, 46.
\bibitem{fermi-0fgl-agn} Abdo, A.A., et al. 2009b, ApJ 700, 597.
\bibitem{dermer2007} Dermer, C.D. 2007, AIP Conf. Proc. 921, 122.
\bibitem{wmap-cat} Wright, E.L., et al. 2009, ApJS 180, 283. 
\bibitem{gtm} Serrano P\'erez-Grovas, A., Schloerb, F.P., Hughes, D., Yun, M. 2006, 
SPIE 6267, 1.
\bibitem{egret-vhe} Thompson, D.J., Bertsch, D.L., O'Neal Jr., R.H. 2005, ApJS 157, 324.
\bibitem{egrcat} Casandjian, JM, Grenier, IA. 2008, A\&A 489, 849.
\bibitem{g2mm-3c454} Carrami\~nana, A., Carrasco, L., Chavushyan, V. et al. 2009,
Fermi Symposium eConf proceedings.
\bibitem{da1968} Galt, J.A., Kennedy, J.E.D. 1968, AJ 73, 135.
\bibitem{misao} http://www.aerith.net/misao/variable/MisV1436.html
\bibitem{qso0133-nir} Carrami\~nana, A., Carrasco, L., Recillas, E., Chavushyan, V. 
2008, ATEL 1874.
\bibitem{qso0133-gam} Takahashi, H., Tosti, G. on behalf of the Fermi-LAT collaboration 
2008, ATEL 1877.
%
\end{thebibliography}

\begin{table*}[t]
\begin{center}
\caption{WMAP sources in the 3EG catalog and 0FGL bright source list.\label{muestra4}}
\begin{tabular}{|l|c|c|c|c|c|c|c|c|}
\hline \textbf{WMAP} & \textbf{3EG / 0FGL} & \textbf{$D$} &
\textbf{Counterpart} &&  
\textbf{WMAP} & \textbf{3EG / 0FGL} & \textbf{$D$} &
\textbf{Counterpart}  \\
\hline J0720--6222 & 3EG J0702--6212& [2] &  \dots &&
         J1510--0546 & 0FGL J1511.2--0536 &  [0]    & PKS 1508--05 \\
\hline J0721+7122 & 0FGL J0722.0+7120 & [0] & PKS 0716+714 &&
          J1512--0904 & 0FGL J1512.7--0905  & [0] & QSO B1510--089 \\ 
          & 3EG J0721+7120 & [0]  &&&         &  3EG J1512-0849 & [0]   &\\
\hline J0738+1743  & 0FGL J0738.2+1738  &  [0]  &  QSO B0735+178 &&  
           J1517--2421 & 0FGL J1517.9--2423  & [0] & QSO B1514--241   \\
             & 3EG J0737+1721 & [0]  &&& & 3EG J1517--2538$^{a}$  &  [1] &  \\
\hline J0808--0750  & 3EG J0812--0646 & [2] & QSO  J0808--0751 && 
           J1608+1027 & 3EG J1608+1055 & [0]   & 4C +10.45 \\
\hline J0813+4817 &  3EG J0808+4844 & [1] & (4C+48.22)     &&  
           J1613+3412  & 3EG J1614+3424 & [1]   & QSO B1611+343 \\
\hline J0825+0311  &  3EG J0828+0508 & [2]  & QSO B0823+033 && 
           J1642+3948 & 0FGL J1641.4+3939 &  [1]  & (4C+39.48)? \\ 
\hline J0831+2411  &  3EG J0829+2413& [0]    & QSO J0830+2411  && 
           J1654+3939 & 0FGL J1653.9+3946 &  [1]   & Mrk 501   \\ 
\hline J0841+7053  &  3EG J0845+7049& [0]  & 0836+710 &&  
           J1703--6214 &  3EG J1659--6251  &[1]  & \\ 
\hline J0854+2006  &  0FGL J0855.4+2009 &  [0]  & OJ 287 && 
           J1736--7934 &  3EG  J1720--7820  & [2] & PKS 1725--795   \\ 
            & 3EG J0853+1941 &  [0] &  &&&&&   \\  
 \hline J0909+0119  &  0FGL J0909.7+0145 & [1]    & PKS 0907+022 &&    
           J1740+5212 & 3EG J1738+5203 &[0]   & QSO B1739+522   \\
\hline J0920+4441 & 0FGL J0921.2+4437  &  [0] &  RGB J0920+446 &&     
           J1800+7827 & 0FGL  J1802.2+7827 & [0]  & S5 1803+78  \\ 
               & 3EG J0917+4427 & [1]  &&&&&&\\
\hline J0957+5527 & 0FGL J0957.6+5522 & [0] &  4C +55.17  &&   
           J1820--6343 & 3EG  J1813--6419 & [1]  &\\  
                            & 3EG J0952+5501& [1] &&&&&&  \\  
\hline J0959+6530 & 3EG J0958+6533 &  [0] & QSO~B0954+65  &&  
           J1848+3223 &  0FGL J1847.8+3223 &  [0]   & TXS 1846+322? \\
\hline J1058+0134 & 0FGL J1057.8+0138 & [0]   & PKS 1055+018  &&    
           J1849+6705 & 0FGL J1849.4+6706 &  [0] & S4 1849+67   \\
\hline J1059--8003  &  0FGL J1100.2--8000  & [0]  & PKS 1057--79 && 
           J1923--2105 & 0FGL J1923.3--2101 & [0] &  PMN J1923--2104 \\  
           &&&&&& 3EG J1921--2015 & [1]   & \\
\hline J1130--1451& 0FGL J1129.8--1443 & [0] & PKS 1127--14  &&   
            J1937--3957  & 3EG J1935--4022 & [2]  & \\
\hline J1147--3811 & 0FGL J1146.7--3808  & [0]  & PKS 1144--379 && 
            J1939--1525 & 3EG J1937--1529 & [0]   & QSO B1939--155   \\
                             &   3EG J1134--1530&  [2]  &&&&&&\\  
\hline J1159+2915 & 0FGL J1159.2+2912 & [0] & 4C +29.45 && 
            J2011--1547 & 3EG  J2020--1545 & [2]   & QSO B2008--159  \\
  & 3EG J1200+2847 & [0] &&&&&&\\
\hline J1223--8306  & 3EG J1249-8330 &  [1]  & &&  
            J2035+1055 &  3EG J2036+1132 & [1]  & QSO~B2032+107  \\
\hline J1229+0203 & 0FGL J1229.1+0202 & [0]  & 3C 273     &&   
            J2056--4716 & 0FGL J2056.1--4715  & [0]    & PMN J2056--4714 \\
                         &    3EG J1229+0210 & [0] &  &&
                         & 3EG J2055--4716 & [0]    &  \\ 
\hline J1246--2547 & 0FGL J1246.6--2544 & [0]  & PKS 1244--255    &&  
            J2143+1741 & 0FGL J2143.2+1741 &  [0] & OX 169  \\ 
\hline J1256--0547 & 0FGL J1256.1--0547 &  [0]   & 3C 279   && 
            J2151--3027 & 3EG  J2158--3023 & [2]  & PKS 2155--304    \\
       & 3EG J1255--0549 & [1] & &&&&& \\ 
\hline  J1310+3222 & 0FGL J1310.6+3220 & [0] & B2 1308+32 &&  
            J2202+4217 & 0FGL  J2202.4+4217  & [0] & Bl Lac  \\ 
       &&&&& & 3EG J2202+4217&  [0] &   \\
\hline  J1316--3337 & 3EG J1314--3431 &[1]  & QSO B1313--333   &&    
            J2203+1723 & 0FGL  J2203.2+1731&  [0]  & PKS 2201+171 \\                        
\hline  J1327+2213 & (3EG J1323+2200)? & [2]   & QSO B1324+224 &&
            J2207--5348  & 0FGL J2207.0--5347 &  [0]   & PKS 2204--54 \\      
\hline  J1337--1257 & 3EG J1339--1419 & [1] & QSO B1335--127  &&  
            J2211+2352 & 3EG J2209+2401&  [0]   & QSO B2209+236  \\ 
\hline  J1354--1041 & 0FGL  J1355.0--1044    & [0]  & PKS 1352--104 &&    
            J2229--0833 & 0FGL J2229.8--0829 &  [0]  & QSO B2227--088   \\
\hline  J1408--0749 & 3EG J1409--0745   & [0] &  QSO B1406--074  &&              
            J2232+1144 & 0FGL  J2232.4+1141  &  [0] & CTA 102 \\
      &&&&&& 3EG J2232+1147 & [0]  &   \\        
\hline  J1419+3822 & 3EG  J1424+3734 & [1]  & QSO B1417+385  &&  
            J2254+1608 & 0FGL  J2254.0+1609 & [0]  & 3C 454.3  \\
       &&&&&        & 3EG J2254+1601 &  [0] & \\
\hline  J1427--4206 & 3EG J1429-4217 & [0]  & QSO B1424--418  &&  
            J2322+4448 &  3EG J2314+4426 & [2]  & GB6 B2319+4429   \\
\hline  J1457--3536 & 0FGL J1457.6--3538 &  [0]   & PKS 1454--354 &&  
            J2327+0937 & 0FGL J2327.3+0947 &  [0]   & GB6 B2325+0923   \\ 
           & 3EG  J1500--3509  &  [0]   & &&&&&   \\
\hline  J1504+1030 & 0FGL J1504.4+1030 &  [0]   & PKS 1502+106   && 
            J2349+3846 &  3EG J2352+3752 & [1] & QSO B2346+385  \\
\hline  J1506--1644 & 3EG J1504--1537  & [1]  & (QSO B1504--1626)  &&   
            J2354+4550 & 3EG J2358+4604 & [1] & 4C 45.51\\
\hline\hline
\end{tabular}
\end{center}
\end{table*}

\end{document}